\documentclass[fluids,article,submit,moreauthors,pdftex,10pt,a4paper]{Definitions/mdpi} 
\firstpage{1} 
\pubvolume{xx}
\issuenum{1}
\articlenumber{5}
\pubyear{2018}
\copyrightyear{2018}
\history{Received: date; Accepted: date; Published: date}

\preto{\abstractkeywords}{\nolinenumbers}

\pdfoutput=1

\Title{Surface waves enhance particle dispersion}

\Author{Mohammad Farazmand$^{\ast}$ \orcidA{} and Themistoklis Sapsis}

\AuthorNames{Mohammad Farazmand and Themistoklis Sapsis}

\address[1]{%
Department of Mechanical Engineering, Massachusetts Institute of Technology\\
77 Massachusetts Ave,  Cambridge, MA 02139-4307, USA
}

\corres{Correspondence: mfaraz@mit.edu; Tel.: +1-617-253-5546}

\abstract{We study the horizontal dispersion of passive tracer particles on the free surface of 
gravity waves in deep water. For random linear waves with the JONSWAP spectrum, the Lagrangian particle trajectories are
computed using an exact nonlinear model known as the John--Sclavounos equation. 
We show that the single-particle dispersion exhibits an unusual super-diffusive behavior. 
In particular, for large times $t$,
the variance of the tracer $\langle |X(t)|^2\rangle$ 
increases as a quadratic function of time, i.e., $\langle |X(t)|^2\rangle\sim t^2$.
This dispersion is markedly faster than Taylor's single-particle dispersion theory
which predicts that the variance of passive tracers grows linearly with time for large $t$. Our results imply that 
the wave motion significantly enhances the dispersion of fluid particles. 
We show that this super-diffusive behavior is 
a result of the long-term correlation of the Lagrangian velocities of fluid parcels on 
the free surface.
}

\usepackage{graphicx,subfigure}
\graphicspath{
{figures/}
}
\usepackage{amsmath,amsfonts,amssymb,amsbsy,amscd,amsgen,mathtools}
\usepackage{amsthm}
\usepackage[]{natbib}
\usepackage{color}
\usepackage{appendix}
\usepackage{comment}
\usepackage{csquotes}

\newcommand{\id}{\mathrm d}

\newcommand{\vc}{\mathbf}

\newcommand{\gr}{\mbox{g}}

\newcommand{\pard}[2]{\frac{\partial #1}{\partial #2}}
\newcommand{\mean}[1]{\langle #1\rangle }

\begin{document}

\section{Introduction}
Water waves cause the material transport of fluid particles on the free surface of the fluid. 
The waves induce a fluid velocity on the free surface which in turn determines the horizontal motion of
fluid particles on the free surface. 
This phenomena has been known since Stokes~\cite{stokes1847} who studied the average velocity of fluid parcels 
transported by a linear monochromatic wave. The resulting displacement is referred to as the \emph{Stokes drift}.

\begin{figure}
\centering
\includegraphics[width=.65\textwidth]{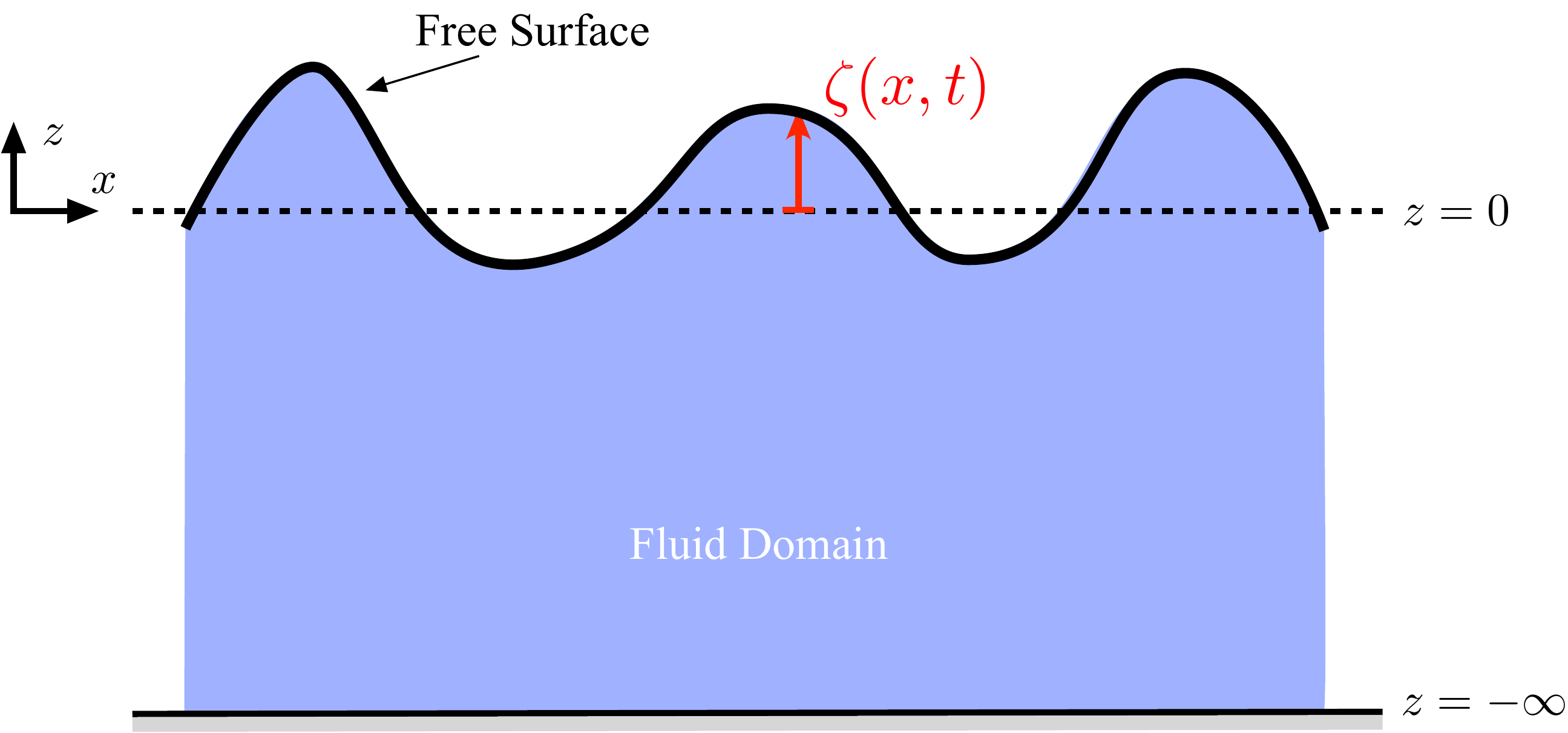}
\caption{Schematic description of the surface waves. The time-dependent free surface $z=\zeta(x,t)$ is a graph over the horizontal coordinate $x$.}
\label{fig:schem}
\end{figure}

More specifically, denote the horizontal trajectory of a fluid particle on the free surface by $x(t;t_0,x_0)$ (or $x(t)$, for short). 
For simplicity, we assume that the waves are unidirectional. The map $x(t;t_0,x_0)$ denotes
the horizontal position of a fluid parcel at time $t$ given its initial position $x_0$ at time $t_0$. 
The surface elevation is assumed to be a graph over the horizontal coordinates and is denoted by $\zeta(x,t)$ (see figure~\ref{fig:schem}). 
Since the fluid parcels on the free surface are constrained to it, their vertical position of the parcel is given by $\zeta(x(t),t)$.
Therefore, knowledge of the horizontal position $x(t)$ of the fluid particles and the free surface elevation $\zeta$
completely determines the position of the particles. 

For a fixed initial condition $x_0$, Stokes' theory~\cite{stokes1847} seeks to determine the average displacement $\mean{x(t)}$ of 
the particles at a later time. Here, $\mean{\cdot}$ denotes the ensemble average over many realizations of the random 
surface waves. In other words, the Stokes' theory is concerned with the first-order statistics of the fluid displacement
on the free surface. The main focus of the present work is the second-order statistics of particle dispersion. More precisely, 
we examine the temporal evolution of the variance $\mean{|X(t)|^2}$ where $X(t)=x(t)-\mean{x(t)}$ denotes the mean-zero displacement of the
fluid particles. 

\subsection{Summary of the main results}\label{sec:mainResults}
As we review in Section~\ref{sec:review}, this second-order statistics has also been studied extensively. 
However, previous studies either assume the velocity field as a linear superposition of 
the velocity induced by linear waves or use perturbation theory to approximate the velocity field. 
Here, however, we obtain 
the fluid particle trajectories $x(t;t_0,x_0)$ from an exact model known as the John--Sclavounos (JS) equation
(see Section~\ref{sec:JS} for a review). 
This exact nonlinear kinematic model was first derived by \citet{john1953} for unidirectional irrotational
water waves. \citet{sclavounos2005} generalized the equations to two-dimensional waves and removed the irrotationality assumption.
\citet{JS16} further analyzed the JS equation discovering its underlying Hamiltonian structure.

Using the JS equation, we compute the temporal evolution of the variance $\mean{|X(t)|^2}$ for fluid particles evolving
from a given initial condition $x(t_0)=x_0$. This puts us in the framework of the single-particle dispersion theory.
Single-particle dispersion was first studied by~\citet{taylor1922} in the context of 
homogeneous, isotropic turbulence (see Section~\ref{sec:taylor}, for a review). Taylor's dispersion
theory predicts that the variance $\mean{|X(t)|^2}$ exhibits a ballistic motion for short times (i.e., $\mean{|X(t)|^2}\sim t^2$)
and a diffusive motion for large times (i.e., $\mean{|X(t)|^2}\sim t$), so that the variance follows the scaling laws
\begin{equation}
\mean{|X(t)|^2} \sim
\begin{cases}
t^2,\quad \mbox{for small}\ t,\\
t,\quad\ \ \mbox{for large}\ t.
\end{cases}
\label{eq:taylor}
\end{equation}

These scaling laws are often assumed in the study of fluid particles dispersed by surface waves.
The main contribution of the present study is to show that Taylor's dispersion laws~\eqref{eq:taylor} may be violated
for particles advected by free-surface waves. More precisely, our results can be summarized as follows.
\begin{enumerate}
\item The variance $\mean{|X(t)|^2}$ of the particles advected by a surface wave
follow the scaling laws
\begin{equation}
\mean{|X(t)|^2} \sim 
\begin{cases}
t^4\quad \mbox{for}\quad t/T_p<1,\\
t^2\quad \mbox{for}\quad t/T_p>1,
\end{cases}
\label{eq:X_scaling}
\end{equation}
where $T_p$ denotes the wave period. This is markedly different from the prediction of Taylor's theory. 
In particular, the long-term evolution of the variance follows a ballistic motion as opposed to a diffusive motion. 

\item Central to Taylor's theory of single-particle dispersion is the autocorrelation function $R(\tau)$ of
the Lagrangian fluid velocities (see Section~\ref{sec:taylor}, for a precise definition of $R$). To arrive at the scaling laws~\eqref{eq:taylor}, Taylor assumes that 
this autocorrelation function decays to zero rapidly enough so that the integral $\int_{0}^{\infty}R(\tau)\id \tau$ exists and its value is finite. 
We show, however, that for particles dispersed by surface waves, the autocorrelation function decays to $1/2$ as 
$\tau\to\infty$. This observation, in turn, explains the unusual scaling~\eqref{eq:X_scaling}. 
\end{enumerate}

\subsection{Earlier studies}\label{sec:review}
Previous studies of particle dispersion on the free surface of a fluid can be 
divided into three general categories: 
i) Passive tracers on a flat free surface,
ii) Linear or nonlinear waves with the induced velocity field modeled based on simplifying assumptions, and
iii) Velocity field derived from satellite altimetry data.

The first category is concerned with
the Lagrangian motion of fluid particles on the \emph{flat} free surface of a fluid in a container.
Although the three-dimensional flow is incompressible, the velocity field restricted to the
surface can be compressible. As a result, the passive tracers can form clusters on a nontrivial
subset of the surface. 
Such no-slip surface flows have been studied experimentally~\cite{Ott1993,Falkovich2006}
and numerically~\cite{Schumacher2002,Schumacher2003,Schumacher2004,Boffetta2004}
with their main emphasis being on the clustering patterns formed by tracers on the free surface.
However, they neglect the effect of waves on the dispersion of the tracers by assuming a flat free surface.

The second category considers the effect of waves on the particle dispersion~\cite{Herterich1982,
Glazman1999,Glazman2000,balk2002,buhler2009,Ferrari2011}. These studies often assume (explicitly or implicitly)
that the longterm particle dispersion is diffusive (Taylor's theory) and aim to approximate the
diffusion tensor.
However, they use a simplified model for the fluid velocity field on the surface. For instance, \citet{Herterich1982}
assume that the Eulerian velocity field is a linear superposition of the velocities obtained 
from linear wave theory. \citet{Glazman1999} also assume such a linear wave theory for
their study of passive tracer advection (also see Refs.~\cite{Glazman2000,balk2002}).

\citet{buhler2009} consider dispersion by random waves in the rotating shallow water framework. 
They go beyond the linear wave theory by using the wave-mean interaction theory to account for 
second-order corrections to the linear velocity field. \citet{Ferrari2011} use a similar approach in the 
framework of rotating Boussinesq equation. Although accounting for the second-order nonlinear effects, 
these studies also assume that the longterm dispersion is diffusive (Taylor's theory). 

The third category derives ocean velocity field from the satellite altimetry data~\cite{Chelton2007,fu2010,Beron-Vera2010,Beron-Vera2013,Olascoaga2013,Beron-Vera2015}. 
These studies are concerned with the large scale mixing in the ocean (on the order of a few kilometers). It is believed that, over such scales, 
the main contribution to mixing comes from the large ocean eddies with negligible contribution from the wave motion.
Nonetheless, the velocity field is derived from the sea surface height measured by altimetry techniques~\cite{Ducet2000}.
In order to relate the sea surface height to the fluid velocity field, one makes the so-called quasi-geostrophic 
assumption, resulting in an approximation of the true velocity field.

In contrast to previous studies, here we use the John--Sclavounos equation to compute the exact Lagrangian trajectory
of fluid parcels on the free surface and thereby examine the validity of the underlying assumptions of Taylor's single-particle dispersion theory.

Before proceeding further, we also refer to the work on \emph{pilot-wave hydrodynamics} which concerns the motion of 
droplets bouncing on the surface of a fluid~\cite{couder2005,Bush2015}. These droplets create a wave when bouncing off the surface
 which in turn guides the motion of the droplet upon subsequent bounces. Clearly, the pilot-wave phenomena is 
 distinct from the dispersion of fluid parcels that belong to the free surface (as considered here) and the two should not be confused. 

\subsection{Outline of the paper}
In Section~\ref{sec:taylor}, we review Taylor's single-particle dispersion theory. 
Section~\ref{sec:JS} reviews the JS equation for the motion of fluid particles on a surface wave.
In Section~\ref{sec:setup}, we describe the set-up of our numerical simulations.
Our numerical results are presented in Section~\ref{sec:numericalResults}.
Finally, we present our concluding remarks in Section \ref{sec:conclude}.

\section{Review of Taylor's single-particle dispersion theory}\label{sec:taylor}
In this section, we briefly review the single-particle dispersion theory
of Taylor~\cite{taylor1922}. This theory is not limited to particle on a free-surface wave
and applies more generally to fluid particles advected by turbulent velocity fields which satisfy the simplifying assumptions 
mentioned below.
Fluid particles move according to the 
ordinary differential equation,
\begin{equation}
\dot{\vc x} = \vc u(\vc x,t),\quad \vc x(t;t_0;\vc x_0)=\vc x_0,
\label{eq:fluid_ode}
\end{equation}
where $\vc x(t;t_0;\vc x_0)\in \mathbb R^d$ denotes the trajectory of a fluid particle starting from the point $\vc x_0$ at the 
initial time $t_0$. The time-dependent vector field $\vc u(\vc x,t)\in\mathbb R^d$ denotes the Eulerian fluid velocity field. 

We also define the Lagrangian velocity $\vc v(t;t_0,\vc x_0) = \vc u(\vc x(t;t_0,\vc x_0),t)$ which measures 
the fluid velocity along the trajectory $\vc x(t;t_0,\vc x_0)$. Integrating~\eqref{eq:fluid_ode} in time, we obtain
\begin{equation}
\vc x(t;t_0,\vc x_0) = \vc x_0 + \int_{t_0}^t \vc v(t';t_0,\vc x_0)\id t'.
\end{equation}

Taylor~\cite{taylor1922} views the Lagrangian velocity $\vc v$ as a stochastic process and seeks to derive the properties of the
resulting stochastic process $\vc x$. In the following, we briefly review Taylor's argument.
For notational convenience, we omit the dependence of the Lagrangian trajectory and velocity on the
initial conditions $(t_0,\vc x_0)$ and simply write $\vc x(t)$ and $\vc v(t)$.

Taylor~\cite{taylor1922} also assumes that the flow is homogeneous and isotropic. These assumptions imply that
the flow can be considered as a one-dimensional motion ($d=1$). Therefore, we omit the vectorization of the 
quantities and denote the fluid parcel's position and velocity by $x(t)$ and $v(t)$, respectively.
We also define the mean-zero position and Lagrangian velocity 
of the particles, 
\begin{equation}
X(t) := x(t)-\mean{x(t)}, \quad V(t) := v(t) - \mean{v(t)},
\end{equation}
where $\mean{\cdot}$ denotes the ensemble average. It is straightforward to show that
\begin{equation}
X(t) = \int_{t_0}^t V(t')\id t'.
\end{equation}

Taylor's theory predicts the scaling~\eqref{eq:taylor} for the variance of the quantity $X(t)$. 
Note that this prediction implies that, for short times,
the particle dispersion is ballistic while, for long times, the particles diffuse as if they are undergoing Brownian motion.
In order to derive expression~\eqref{eq:taylor}, we consider the time derivative of the variance of $X(t)$,
\begin{equation}
\frac{\partial}{\partial t} \mean{|X(t)|^2} = 2\mean{X(t)V(t)} = 2\int_{t_0}^t \mean{ V(t)V(t')}\id t'=2\int_0^t \mean{V(t)V(t-\tau)}\id \tau, 
\label{eq:varX_dt}
\end{equation}
where for the last identity we used the change of variables $\tau=t-t'$ and assumed $t_0=0$ (without loss of generality). 
Assuming that $V$ is a stationary process, the covariance $\mean{V(t)V(t-\tau)}$ only
depends on $\tau$. Therefore, the autocorrelation function $R(\tau) := \mean{V(t)V(t-\tau)}/\mean{|V(t)|^2}$ only depends on 
the delay parameter $\tau$. Integrating equation~\eqref{eq:varX_dt} in time,
we obtain
\begin{align}
\mean{|X(t)|^2}  & = 2\mean{|V(t)|^2} \int_0^t\int_0^{t'} R(\tau)\id \tau\id t'\nonumber\\
& = 2\mean{|V(t)|^2}\int_0^t(t-\tau) R(\tau)\id \tau.
\label{eq:varX_R}
\end{align}

We note that, for homogeneous and isotropic flows with stationary Lagrangian velocities $v(t)$, equation~\eqref{eq:varX_R} is exact. 
In order to arrive at the scaling laws~\eqref{eq:taylor}, Taylor~\cite{taylor1922} makes further simplifying assumptions.
In particular, he assumes that, for small $t$, the autocorrelation
function is constant. As a result, for small $t$, we obtain $\mean{|X(t)|^2}\sim 2\mean{V^2}\int_0^t(t-\tau)\id \tau=\mean{V^2}t^2$
(note that the variance $\mean{V^2}$ is time-independent since we assumed the Lagrangian velocity is a stationary stochastic process).
Furthermore, assume that $\int_0^\infty R(\tau)\id \tau$ and $\int_0^\infty \tau R(\tau)\id \tau$ 
exist and are finite. Then, for large $t$, we have $\mean{|X(t)|^2}\sim 2Dt$ where $D=\mean{V^2}\int_0^\infty R(\tau)\id \tau$.

To summarize, for the scaling laws~\eqref{eq:taylor} to hold, the stochastic Lagrangian velocities $v(t)$ must be stationary. Furthermore, 
the autocorrelation function $R(\tau)$ must decay to zero fast enough so that the integrals $\int_0^\infty R(\tau)\id \tau$ and $\int_0^\infty \tau R(\tau)\id \tau$
are finite. In Section~\ref{sec:numericalResults}, we show that some of these assumptions do not hold generally for particles advected by 
random surface waves. Our results are obtained by numerically integrating an exact model of the fluid trajectories on the free surface. 
We introduce this model in the next section. 

\section{John--Sclavounos equation}\label{sec:JS}
The John--Sclavounos (JS)~\cite{john1953,sclavounos2005} equation describes the 
horizontal motion of fluid particles on a free surface $\zeta$. Lets denote the horizontal coordinate 
by $x$ and denote the vertical coordinate (corresponding to the depth) by $z$. 
We assume that the free surface is a graph over the horizontal coordinate $x$ so that on the free surface $z=\zeta(x,t)$ (see figure~\ref{fig:schem}).
The position of the particles 
on the free surface at time $t$ is then given by $(x(t),\zeta(x(t),t))$. The Eulerian velocity field inside the fluid 
is denoted by $\vc u(x,z,t)$. The fluid velocity $\vc u$ and the free surface $\zeta(x,t)$ satisfy the water wave equations. 

With this notation, the JS equation reads
\begin{equation}
\ddot x = -\frac{\zeta_{,xx}\dot x^2+2\zeta_{,xt}\dot x+(\gr+\zeta_{,tt})}{1+\zeta_{,x}^2}\zeta_{,x},
\label{eq:JS}
\end{equation}
where $\zeta_{,x}$ is shorthand notation for the partial derivative $\partial \zeta/\partial x$ and similarly for partial derivatives with respect to time.
Supplying equation~\eqref{eq:JS} with the appropriate initial conditions $(x(0), \dot x(0))$ and integrating in time, 
the horizontal motion of the particles on the free surface can be computed. 
The JS equation is quite a remarkable result. It implies that if the surface elevation $\zeta$ is known, then the 
motion of the particles on the free surface can be deduced without knowing the full fluid velocity field $\vc u(x,z,t)$.

Denoting the horizontal Lagrangian velocity of a fluid parcel by $v=\dot x$, we write the JS equation as a set of first-order differential equations
\begin{equation}
\dot x = v, \quad \dot v = -\frac{Q(x,v,t)}{1+\zeta_{,x}^2}\zeta_{,x}, 
\label{eq:JS-1order}
\end{equation}
where \begin{equation}
Q(x,v,t):=\zeta_{,xx}v^2+2\zeta_{,xt}v+(\gr+\zeta_{,tt}) = \frac{\id^2\ }{\id t^2}\zeta (x(t),t)+\gr.
\label{eq:Q}
\end{equation}
Fedele et al.~\cite{JS16} showed that the JS equations~\eqref{eq:JS-1order} have a Hamiltonian structure
which in the 1D case is given by
\begin{equation}
\dot x = \pard{H}{p},\quad \dot p = -\pard{H}{x},
\end{equation}
where the Hamiltonian $H$ reads
\begin{equation}
H(x,p,t)= \frac12 \frac{(p-\zeta_{,t}\zeta_{,x})^2}{1+\zeta_{,x}^2}+\gr\zeta -\frac12 \zeta_{,t}^2,
\end{equation}
and the generalized momentum $p$ is given by
$p = (1+\zeta_{,x}^2) v + \zeta_{,t}\zeta_{,x}$.

In the time-independent case, where $\zeta_{,t}\equiv 0$, the Hamiltonian is a conserved quantity. 
In other words, the quantity,
\begin{equation}
H_0 = \frac12 \frac{p^2}{1+\zeta_{,x}^2}+\gr\zeta,
\end{equation}
is invariant along the trajectories of the JS equation. However, in the realistic situation where the free surface elevation is time-dependent, 
the energy $H$ is no longer conserved and complex particle motion is possible.

The derivation of~\citet{JS16} also shows that the JS equation holds more generally for any particle constrained 
on a surface $\zeta(x,t)$ and moving under the gravitational force. In particular, they show that the JS equation
can be derived without making use of the continuity equation. 

In the following, we do not leverage the Hamiltonian structure of the JS equation. Instead, for a given 
surface elevation $\zeta(x,t)$, we numerically integrate the JS equation~\eqref{eq:JS-1order}
and compute the resulting particle dispersion properties. Although here we focus on unidirectional waves, 
the JS equation is applicable to two-dimensional waves (see Refs.~\cite{sclavounos2005,JS16}) and therefore
our results can be generalized in a straightforward fashion.

We point out that a model similar to the JS equation was proposed by \citet{Janssen2016} (see their equation (20)). 
That model however is inaccurate since it neglects the denominator in equation~\eqref{eq:JS}.

\section{Set-up}\label{sec:setup}
\subsection{Irregular wave field}
We consider random surface waves in deep water consisting of a finite sum of plane waves, 
\begin{equation}
\zeta(x,t) = \sum_{i=1}^n a_i \cos (k_i x-\omega_i t + \phi_i),
\label{eq:random_wave}
\end{equation}
where $k_i$ is the wave number, $\omega_i$ is the wave frequency satisfying the linear dispersion relation $\omega_i = \sqrt{g k_i}$.
The random phases $\phi_i$ are uniformly distributed over the interval $[0,2\pi]$.

We consider waves that follow the JONSWAP (Joint North Sea Wave Project) spectrum~\cite{hasselmann1973}
\begin{equation}
S(\omega) =  \frac{\alpha \gr^2}{\omega^5} 
\exp\left[-\frac{5}{4}\left(\frac{\omega_p}{\omega}\right)^4\right] 
\gamma^{\exp\left[-\frac{(\omega/\omega_p - 1)^2}{2\beta^2}\right]},
\label{eq:jonswap}
\end{equation}
where $\gamma = 3.3$ is the enhancement factor and $\omega_p$ is the
angular frequency corresponding to the peak of the spectrum.
The standard deviation $\beta=0.07$ for $\omega\leq \omega_p$
and $\beta=0.09$ for $\omega> \omega_p$. The amplitude $\alpha$ will be specified later according to 
the desired wave steepness. 

The wave amplitudes in~\eqref{eq:random_wave} are set to $a_i = \sqrt{2S(\omega_i)\Delta\omega}$
so that the random wave field $\zeta$ has the JONSWAP spectrum~\eqref{eq:jonswap}.
This yields $\sigma^2:=\langle \zeta ^2 \rangle = \frac12 \sum_{i=1}^{n} a_i^2 = \sum_{i=1}^{n} S(\omega_i)\Delta \omega$,
where $\sigma$ is the standard deviation of the wave elevation $\zeta$.
The significant wave height $H_s$ is defined as four times this standard deviation, $H_s=4\sigma$.
Following~\cite{Onorato13}, we define the average wave steepness as $\epsilon = H_sk_p/2$.

\subsection{Initial conditions}\label{sec:v0}
The JS equation~\eqref{eq:JS-1order} is a two-dimensional system of first-order differential equations. In order to numerically integrate these equations, 
we need to supply them with the appropriate initial conditions $(x(0),v(0))=(x_0,v_0)$. Since the surface waves~\eqref{eq:random_wave}
are stochastically homogeneous, the choice of the initial position $x_0$ does not alter the final ensemble averaged quantities. Therefore, 
we simply set $x_0=0$.

The initial velocity $v_0$ should be consistent with the Eulerian velocity induced by 
the wave motion~\cite{sclavounos2005}. Since this Eulerian velocity is unknown (unless one solves the full water wave equations), 
we have to make an assumption for $v_0$. 
Here, we set the initial particle velocity $v_0$ to coincide with the induced velocity of a 
linear wave. The velocity potential of a monochromatic linear wave at the surface is 
$\phi = (a\gr /\omega) \sin (kx-\omega t)$. This implies the horizontal
velocity $v = \partial \phi/\partial x= (agk/\omega)\cos(kx-\omega t)$. 
Since the surface is a superposition of linear waves, each component of the
random wave field contributes differently to the horizontal particle velocity.
For simplicity, we assume that the main contribution comes
from the wave corresponding to the peak of the spectrum. Therefore, 
we set $v_0 = agk_p/\omega_p$. Furthermore, the wave amplitude $a$ 
satisfies $\langle \zeta^2\rangle = \frac12 a^2$. This implies
\begin{equation}
v_0 = \frac{\sqrt{2}\sigma g k_p}{\omega_p},
\label{eq:v0}
\end{equation}
where $\sigma = \langle \zeta^2\rangle^\frac12$ is the standard deviation of 
the wave elevation. We have checked numerically that the resulting ensemble averaged Lagrangian velocity
is time-independent and $\mean{v(t)}=v_0$; thereby confirming that the choice of the initial velocity does not affect the
longterm ensemble behavior of the particles. Furthermore, we have perturbed this initial velocity $v_0$ and observed no 
significant change in the results reported in section~\ref{sec:numericalResults}.

Although the initial velocity $v_0$ is deterministic, $v(t)$ is stochastic for $t>0$. 
This is because $v$ satisfies the JS equation~\eqref{eq:JS-1order}
whose right-hand-side inherits the stochasticity of the surface elevation~\eqref{eq:random_wave}.

\subsection{Non-dimensional variables}\label{sec:nondim}
In order to nondimensionalize the variables,
we rescale space and time as $x\mapsto x/\lambda_p$ and $t\mapsto t/T_p$, respectively.
The length scale $\lambda_p$ is the wavelength corresponding to the peak of the JONSWAP spectrum~\eqref{eq:jonswap}
and $T_p=2\pi/\omega_p$ is the associated wave period. 

The gravitational constant in non-dimensional form becomes $g\mapsto g T_p^2/\lambda_p$.
The linear dispersion relation $\omega_p^2 = gk_p$ implies $gT_p^2/\lambda_p=2\pi$, i.e., 
the non-dimensional gravitational constant is $\gr =2\pi$. The initial velocity~\eqref{eq:v0}
in terms of non-dimensional variables becomes $v_0 = \epsilon\sqrt 2/2$.


\section{Numerical results}\label{sec:numericalResults}
\begin{figure}
	\centering
	\includegraphics[width=.6\textwidth]{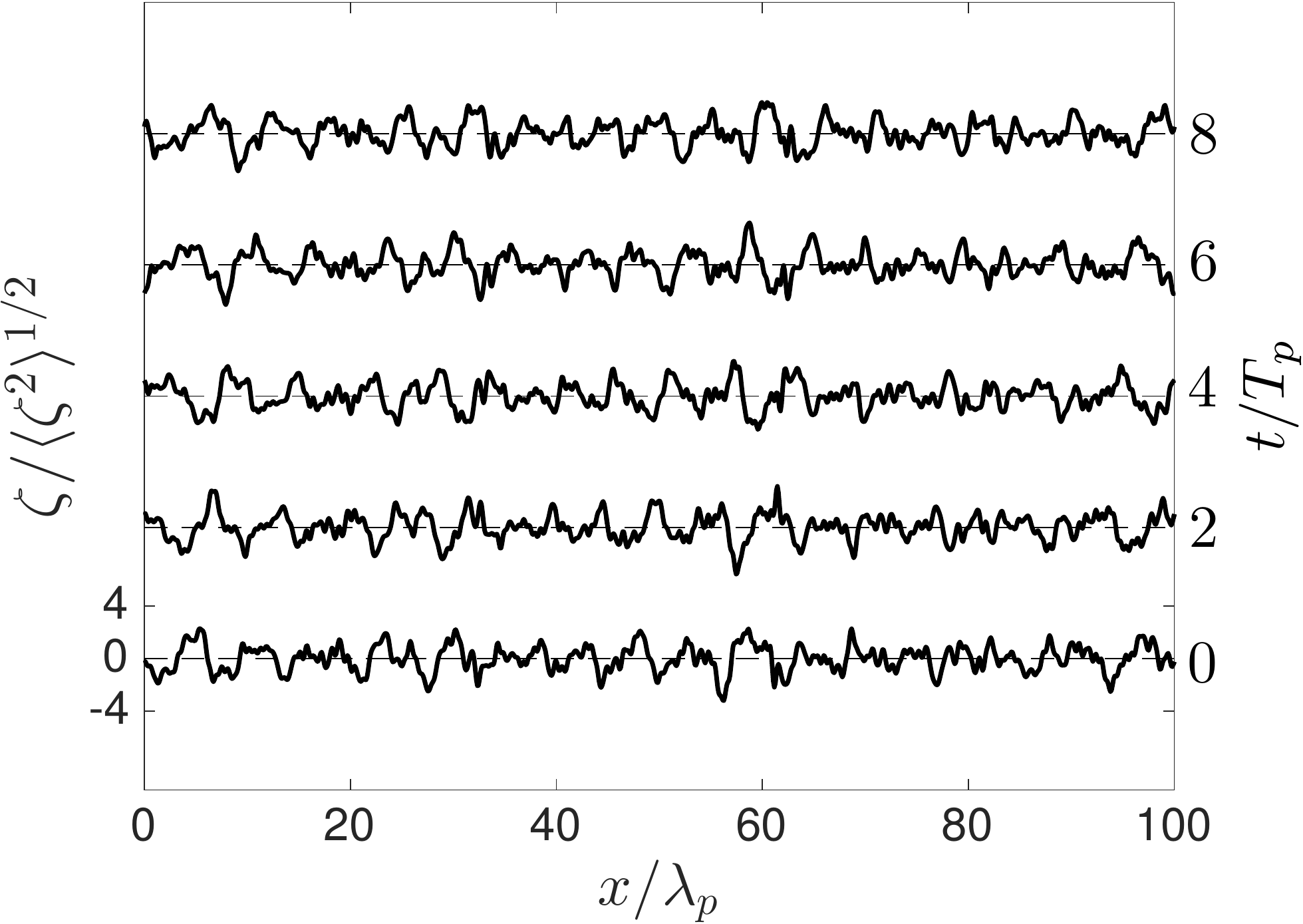}
	\caption{The space-time evolution of the random wave field~\eqref{eq:random_wave} with steepness $\epsilon=0.05$.
		The horizontal dashed lines mark $\zeta=0$.}
	\label{fig:wave}
\end{figure}

In this section, we present the results obtained from numerically integrating the JS equation~\eqref{eq:JS-1order}. 
All results are reported in the non-dimensional variables discussed in section~\ref{sec:nondim}.
First, we generate the random wave fields from~\eqref{eq:random_wave} with $n=200$ and the JONSWAP spectrum~\eqref{eq:jonswap}. 
The computational domain is $x\in [0,200\pi]$ with periodic boundary conditions. The domain is large enough to avoid finite-box effects. 
The parameter $\alpha$ in the JONSWAP spectrum~\eqref{eq:jonswap} is set to $\alpha = 210(\epsilon/2)^2$ which 
guarantees that the resulting waves have average steepness $\epsilon$.
We report our results for three wave steepness values $\epsilon=0.05$, $0.075$ and $0.1$ which are below the threshold for 
breaking waves (Recall that JS equation is not valid for breaking waves). Figure~\ref{fig:wave} shows a realization of 
the random wave field with steepness $\epsilon=0.05$.

Given a wave field $\zeta$, we integrate the JS equation~\eqref{eq:JS-1order} with 
the embedded Runge--Kutta scheme \texttt{RK5(4)} of MATLAB~\cite{RK45}. The initial conditions are $x(0)=x_0=0$ and
$v(0)=v_0$ with the initial velocity $v_0$ discussed in section~\ref{sec:v0} (see equation~\eqref{eq:v0}).
For each steepness $\epsilon$, we generate $50,000$ random waves. For each realization of the
waves, we compute the particle trajectories and estimate the 
ensemble averages $\mean{\cdot}$ from these 50,000 trajectories. 

\begin{figure}
\centering
\subfigure[]{\includegraphics[width=.45\textwidth]{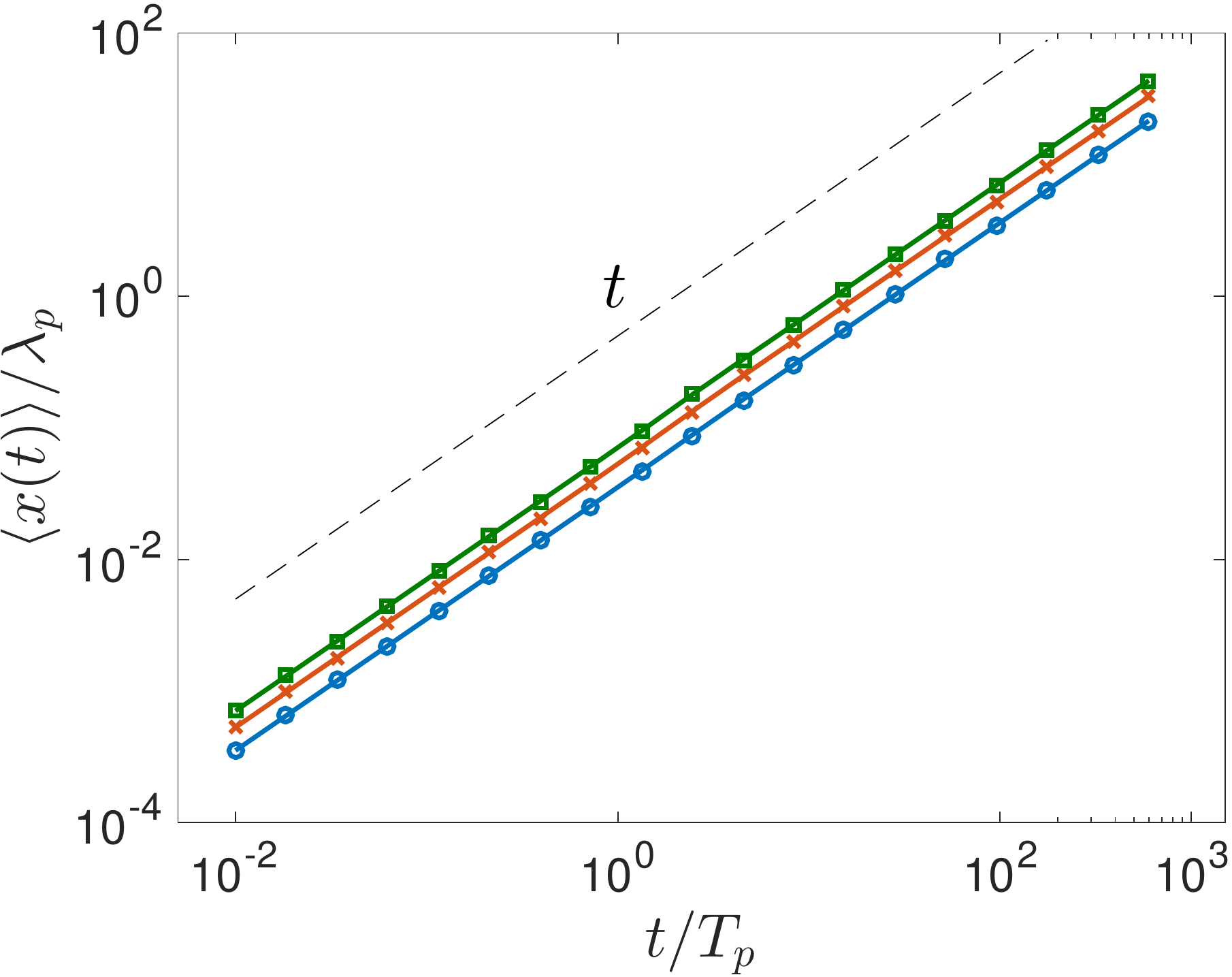}}\hspace{.09\textwidth}
\subfigure[]{\includegraphics[width=.45\textwidth]{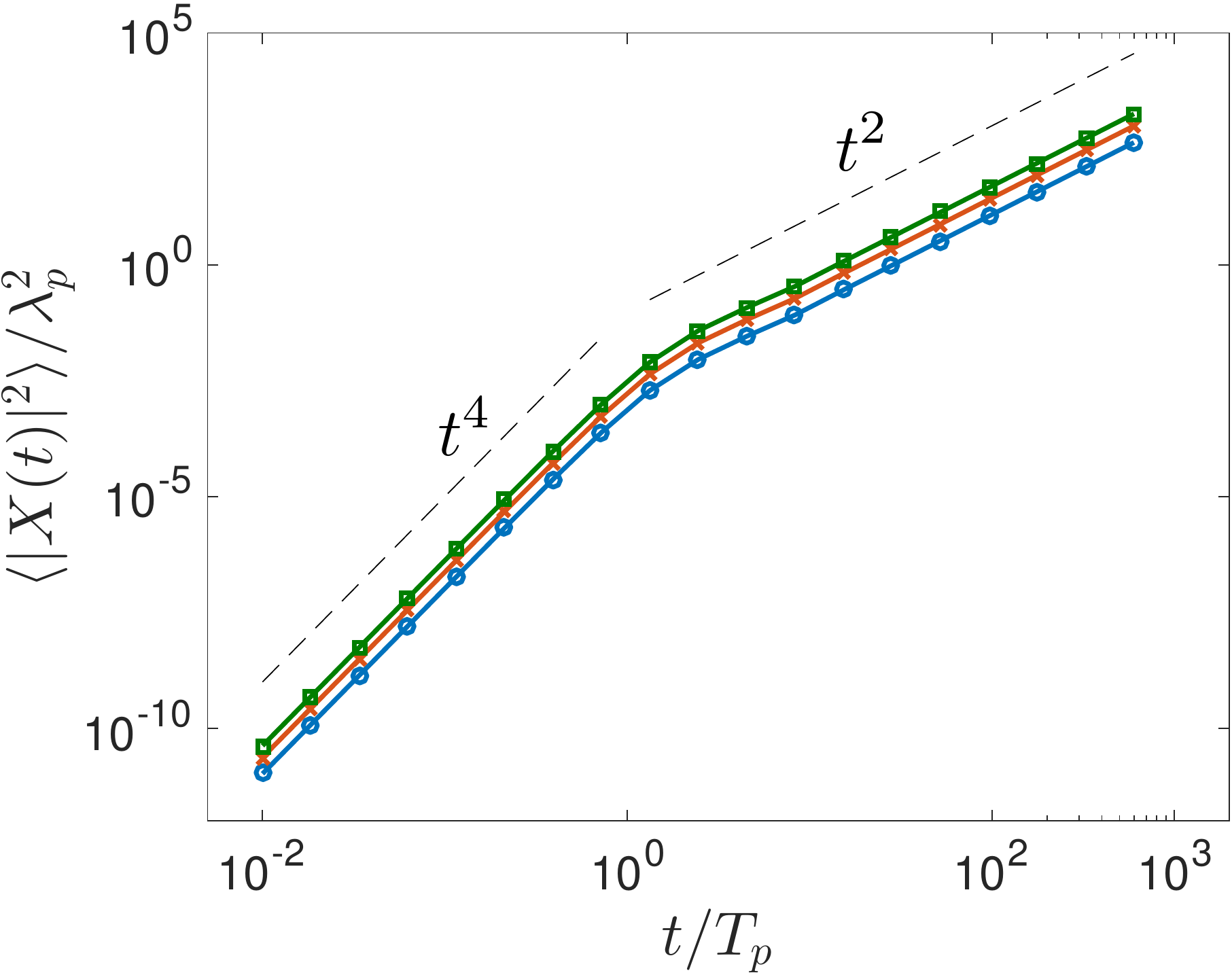}}
\caption{The average drift $\langle x(t)\rangle$ (a) and the variance $\langle |X(t)|^2\rangle$ (b) of the particles. 
Different colors (symbols) correspond to wave steepness $\epsilon=0.05$ (blue circles), 
$\epsilon=0.075$ (orange crosses) and $\epsilon=0.1$ (green squares). The dashed lines mark the indicated slopes.}
\label{fig:x_mean_var}
\end{figure}

First, we examine the average drift $\mean{x(t)}$ as shown in figure~\ref{fig:x_mean_var}(a). 
The average drift grows linearly with time $t$ which agrees with Stokes' prediction~\cite{stokes1847,vandenBremer2017}.
However, the variance $\mean{|X(t)|^2}$ of the particle positions exhibits a surprising behavior.
Recall that $X(t) = x(t)-\mean{x(t)}$ is the mean-zero position of the particles.
Figure~\ref{fig:x_mean_var}(b) shows this variance as a function of time.
For short times $\langle |X(t)|^2\rangle$ grows with
the power law scaling $t^4$. After roughly one wave period, the growth of the variance 
changes and scales as $t^2$. These two regimes result in the scaling law~\eqref{eq:X_scaling}.

\begin{figure}
\centering
\subfigure[]{\includegraphics[width=.45\textwidth]{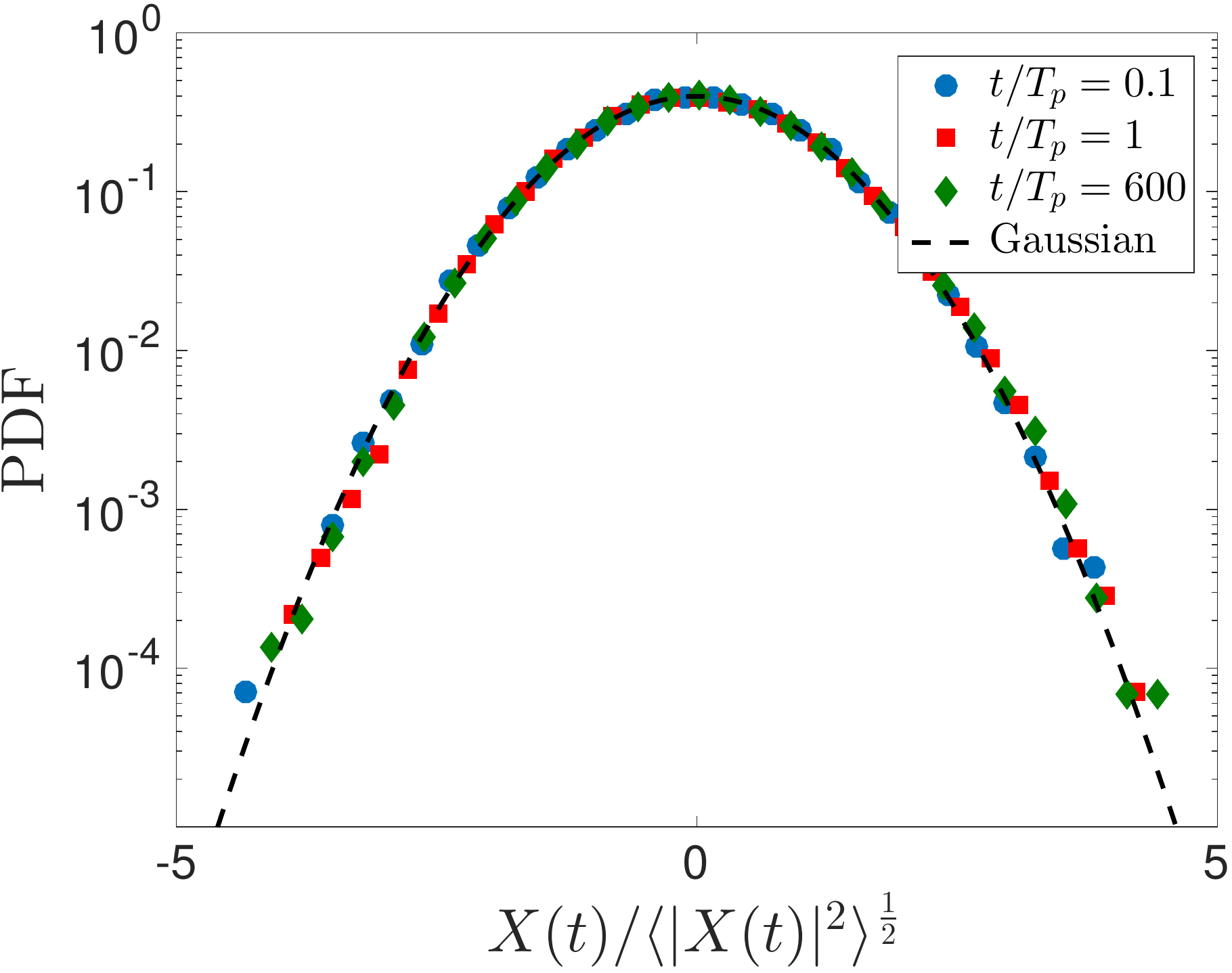}}\hspace{.09\textwidth}
\subfigure[]{\includegraphics[width=.45\textwidth]{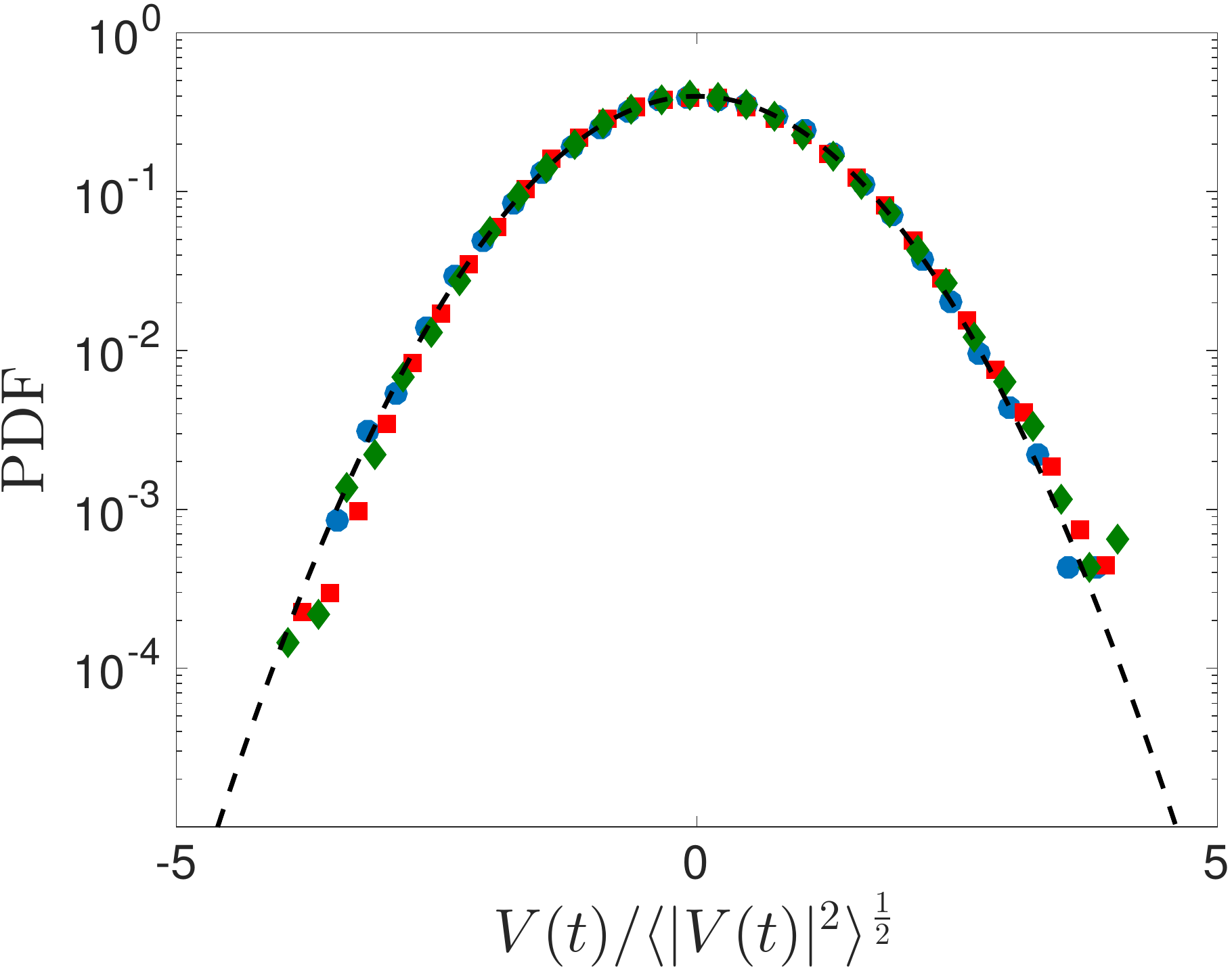}}
\caption{Probability density of the particle position $X(t)=x(t)-\mean{x(t)}$ and velocity $V(t)=v(t)-\mean{v(t)}$ normalized by their standard deviations. 
The PDF is shown at three times with the particle dispersion corresponding to the wave steepness $\epsilon=0.1$.
The dashed line marks the standard normal distribution.}
\label{fig:pdf}
\end{figure}

Note that this behavior is very different from Taylor dispersion theory~\eqref{eq:taylor} in the absence of waves. 
Taylor dispersion predicts a scaling $t^2$ for short times while, in the presence of waves, we 
observe the scaling $t^4$. For large times, Taylor dispersion predicts a Brownian-type 
diffusion where the variance increases linearly with time while, in the presence of waves, we 
observe a ballistic dispersion with $\mean{|X(t)|^2}\propto t^2$. 

This super-diffusive asymptotic behavior has been 
reported in homogeneous, isotropic turbulence where $\mean{|X(t)|^2}\propto t^\gamma$
with $\gamma>1$~\cite{klafter87, warhaft2000, del-Castillo2005}. 
In turbulence, the departure from Taylor dispersion theory is often associated with 
intermittency which manifests itself as heavy tails in the distributions of the particle positions 
and velocities. However, for particles dispersion by waves, we do not observe such heavy-tail statistics. 
In fact, as figure~\ref{fig:pdf} shows, the PDF of particle positions and velocities are 
Gaussian. Note that the initial conditions $(x_0,v_0)$ are deterministic and therefore 
their initial distributions are delta functions. However, as shown in figure~\ref{fig:pdf}, 
they rapidly converge to a Gaussian for $t>0$ (e.g., see the circles marking the PDFs at time $t/T_p=0.1$).
As we show in Appendix~\ref{app:gaussian}, these Gaussian distributions can be deduced from the
JS equations and a central limit theorem.

\begin{figure}
\centering
\includegraphics[width=.6\textwidth]{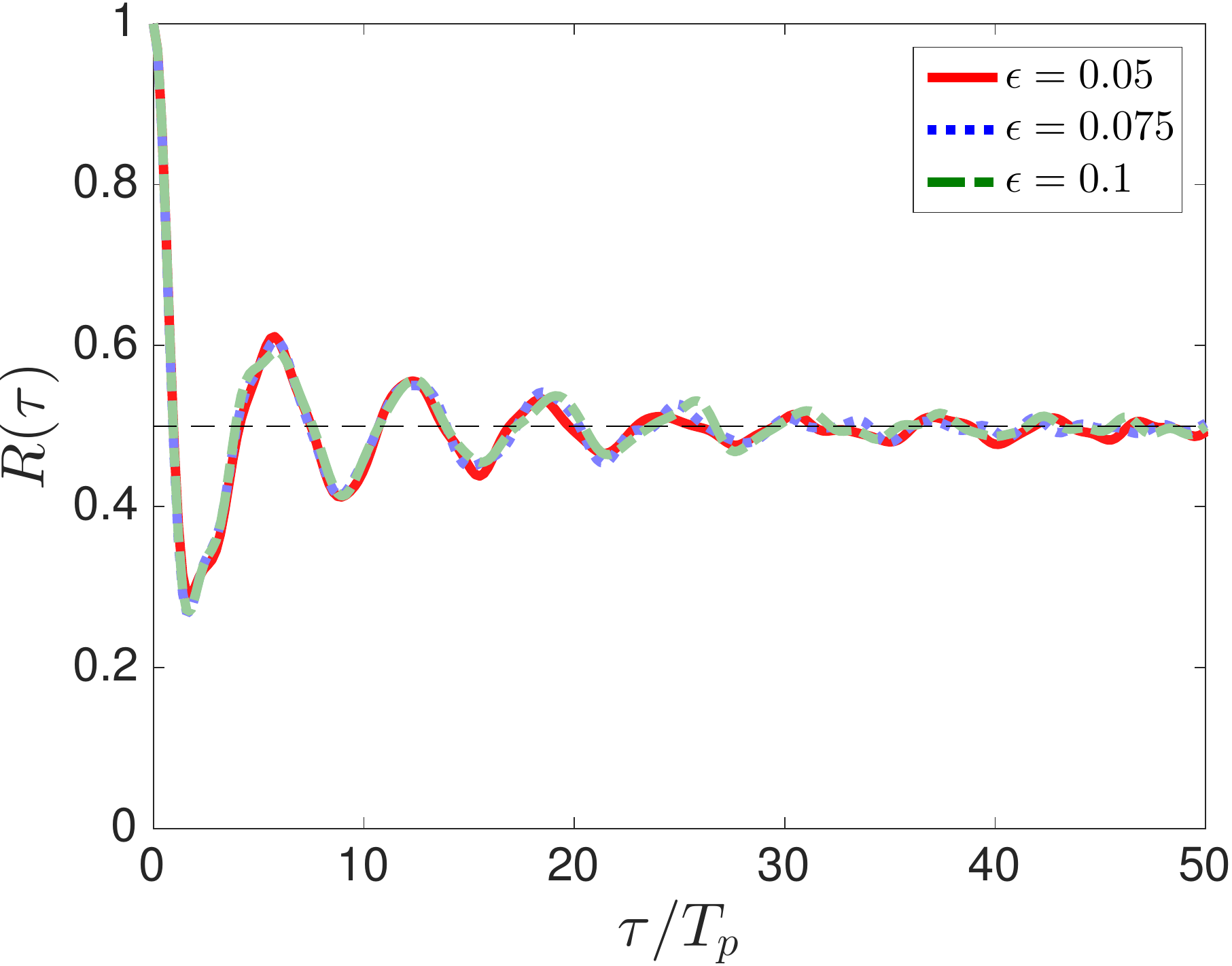}
\caption{The autocorrelation function $R(\tau)=\mean{V(t)V(t-\tau)}/\mean{|V(t)|^2}$ 
for three different wave steepnesses $\epsilon$. The black dashed line marks $R=\frac12$.}
\label{fig:autoCorr}
\end{figure}

The unusual scaling~\eqref{eq:X_scaling} observed here can be explained by examining the autocorrelation function
$R(\tau) = \mean{V(t)V(t-\tau)} /\mean{|V(t)|^2}$. For $t$ large enough, this function is independent of $t$ and
only depends on the delay $\tau$. Figure~\ref{fig:autoCorr} shows $R(\tau)$ computed for $t=100T_p$.
One important feature of this autocorrelation function is that it does not decay to zero for large $\tau$. Instead, 
it decays to $R(\tau)=1/2$, indicating that the Lagrangian particle velocities on the free surface remain correlated indefinitely.

\begin{figure}
	\centering
	\includegraphics[width=.6\textwidth]{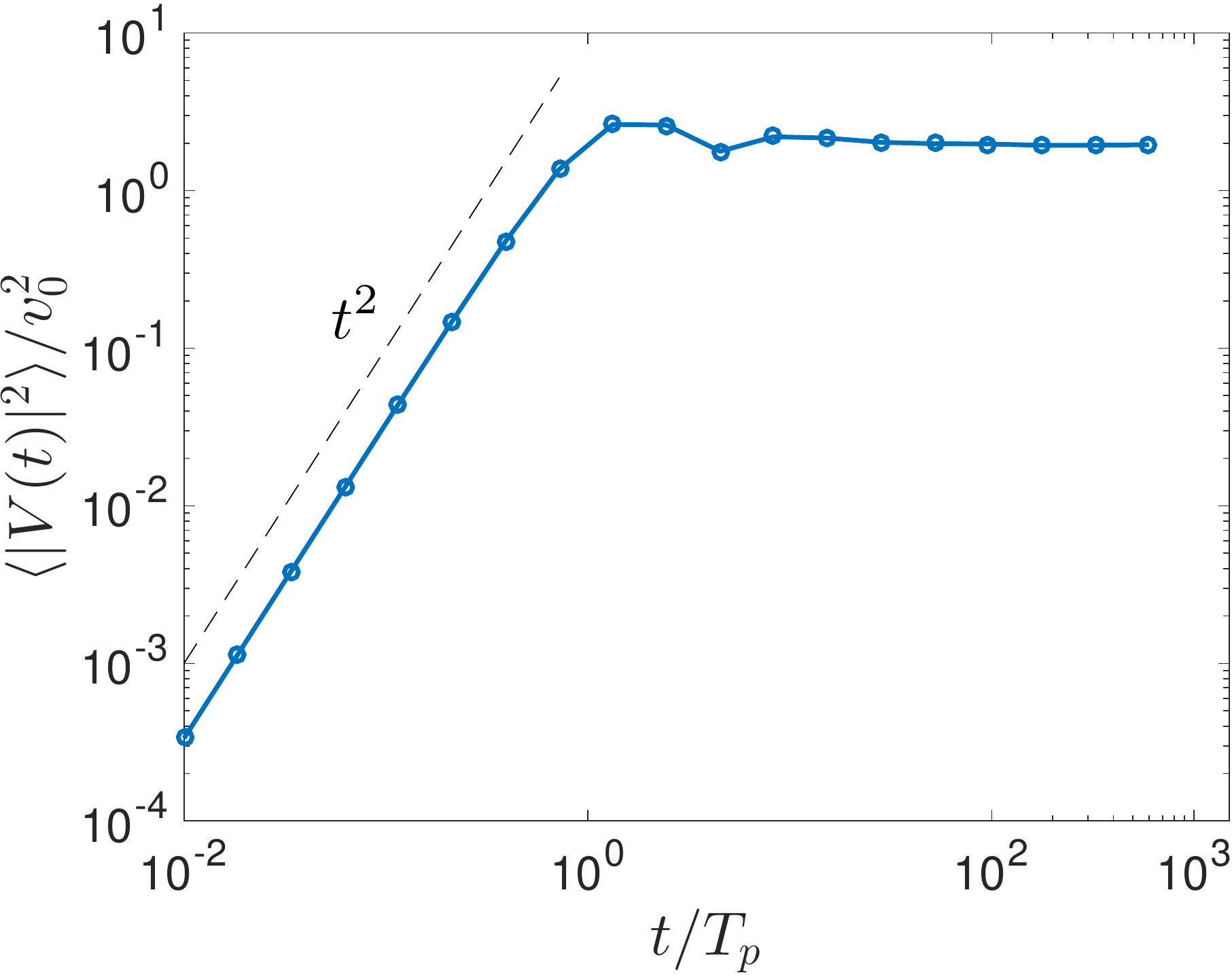}
	\caption{The variance of the Lagrangian velocities $v(t)$ normalized by $v_0^2$ where $v_0=\epsilon\sqrt{2}/2$ is 
		the non-dimensional initial initial velocity discussed in section~\ref{sec:v0}. This plot is identical for all three
		wave steepnesses $\epsilon=0.05$, $0.075$ and $0.1$.}
	\label{fig:var_vt}
\end{figure}

This correlation explain the ballistic motion of particles observed in figure~\ref{fig:x_mean_var}(b) 
for large $t$, i.e., $\mean{|X(t)|^2}\propto t^2$. Recall equation~\eqref{eq:varX_R} that relates the 
variance of the particles to the autocorrelation function. For large $\tau$, the autocorrelation tends to $1/2$
and therefor the integral $\int_0^t (t-\tau)R(\tau)\id\tau$ scales as $\int (t-\tau)/2\id \tau\sim t^2/4$ as $t\to\infty$.
Furthermore, the variance $\mean{|V(t)|^2}$ is constant for large $t$ (see figure~\ref{fig:var_vt}). 
These numerical observations, together with equation~\eqref{eq:varX_R}, imply that 
$\mean{|X(t)|^2}\sim t^2$.

Finally, we turn our attention to the short-term behavior $\mean{|X(t)|^2}\sim t^4$ reported in figure~\ref{fig:x_mean_var}(b). 
Figure~\ref{fig:var_vt} shows that, for $t<T_p$, the variance $\mean{|V(t)|^2}$ increases quadratically in time. This together with equation~\eqref{eq:varX_R}
implies that the variance $\mean{|X(t)|^2}$ of the fluid parcels should in fact increase as $t^4$.

\section{Conclusions}\label{sec:conclude}
Here, we investigated the nonlinear dispersion of fluid parcels on the
free surface of a random wave in deep ocean. The free surface is 
assumed to be a superposition of linear waves with random phases and the JONSWAP spectrum.
However, the fluid particle trajectories are computed using an 
exact nonlinear model known as the John--Sclavounos (JS) equation~\cite{john1953,sclavounos2005,JS16}.

Our main finding is the breakdown of Taylor's dispersion theory~\cite{taylor1922}. 
In particular, for large times, the particles disperse ballistically as opposed to diffusively.
More precisely, the variance $\mean{|X(t)|^2}$ of the distribution of the particles increases quadratically in time 
for large $t$ so that $\mean{|X(t)|^2}\sim t^2$. For short times, the variance is proportional to $t^4$
giving rise to the scaling laws~\eqref{eq:X_scaling}.

We showed that this unusual scaling law is 
a consequence of longterm correlation of Lagrangian velocities of the fluid particles. 
This is a clear violation of Taylor's assumption that these correlations decay 
to zero asymptotically. Clearly, our results have profound implications for modeling
the dispersion of fluid particles (and pollutants) on the ocean surface. 

Since the JS equations are valid in two dimensions, extending our results to two-dimensional waves is 
straightforward and will be presented in future work. Furthermore, future work will investigate the single-particle 
dispersion on the surface of (weakly) nonlinear waves.
This can be accomplished, for instance, by the one-way coupling of the JS equation to an
envelop equation for the free surface (e.g., the nonlinear Schr\"odinger equation). 

\acknowledgments{This work has been supported through the ARO MURI W911NF-17-1-0306 and 
the ONR grant N00014-15-1-2381.
M. F. acknowledges fruitful discussions with John Bush (Department of Mathematics, MIT).}

\appendixsections{multiple}
\appendix
\section{Gaussian distribution of the particle position and velocity}\label{app:gaussian}
In this section, we show that the Gaussian behavior observed in figure~\ref{fig:pdf} can 
be deduced directly from the JS equation.
Recall that the JS equations are valid for relatively small-amplitude waves so that wave breaking does not take place. 
More precisely, the wave surface $\zeta(x,t)$ is $\mathcal O(\epsilon)$ where $0<\epsilon\ll 1$ is the wave steepness. 
Therefore, to the leading order, the JS equation~\eqref{eq:JS} reads
\begin{equation}
\ddot x = -\gr\, \zeta_{,x}(x,t) = \gr \sum_{i=1}^n a_ik_i\sin (k_ix-\omega_i t+\phi_i),
\label{eq:vdot_O1}
\end{equation}
where we used the random wave expression~\eqref{eq:random_wave} for the second identity.
The solution $x(t;t_0,x_0)$ has a $\mathcal O(1/\sqrt{n})$ weak dependence on each
of the random phases $\phi_i$. Therefore,
the terms $\sin(k_ix-\omega_i t+\phi_i)$ are independent, identically distributed (i.i.d.) random variables
up to an error of order $\mathcal O(1/\sqrt{n})$.

The terms $a_i k_i\sin (k_ix-\omega_i t+\phi_i)$ are also independent but they are not identically
distributed due to the wights $a_ik_i$. Applying the Lyapunov Central Limit Theorem (CLT)~\cite{sinai1992} to the sum in equation~\ref{eq:vdot_O1}, we conclude
that $\ddot x$ is a Gaussian process with zero mean. 
Furthermore, this implies that $\dot x$ is Gaussian with constant mean and
$x$ is a Gaussian process whose mean increases linearly with time, i.e. $\mean{x(t)}\sim t$ (see figure~\ref{fig:x_mean_var}(a)).

\reftitle{References}

\end{document}